\documentclass[aps,groupedaddress,twocolumn]{revtex4-1}
\usepackage[dvips]{graphicx}
\usepackage{amssymb}

\usepackage{color}
\usepackage{url}

\begin{document}
\title{Crystal growth, characterization and advanced study of the noncentrosymmetric superconductor Mo$_3$Al$_2$C}

\author{N. D. Zhigadlo}
\affiliation{Department of Chemistry and Biochemistry, University of Bern, 3012 Bern, Switzerland}
\email{nzhigadlo@gmail.com}

\author{D. Logvinovich}
\affiliation{Laboratory of Crystallography, Department of Materials, ETH Zurich, 8093 Zurich, Switzerland }

\author{R. S. Gonnelli}
\affiliation{Department of Applied Science and Technology, Politecnico di Torino, 10129
Torino, Italy }

\author{V. A. Stepanov}
\affiliation{Division of Solid State Physics, P. N. Lebedev Physical Institute, Russian Academy of Sciences, Moscow, Russia}

\author{D. Daghero}
\affiliation{Department of Applied Science and Technology, Politecnico di Torino, 10129
Torino, Italy }
\email{dario.daghero@polito.it}

\date{\today}
\begin{abstract}
We report on the first successful growth of single crystals of the noncentrosymmetric superconductor Mo$_3$Al$_2$C obtained by means of a cubic-anvil, high-pressure and high-temperature technique. Composition, structure, and normal-state transport properties of the crystals were studied by means of X-ray diffraction, energy-dispersive X-ray spectroscopy, magnetic susceptibility and resistivity measurements as a function of temperature. Variations in critical temperature ($T_c$) between 8.6 and 9.3 K were observed, probably due to the slightly different carbon stoichiometry of the samples. Single-crystal X-ray refinement confirmed the high structural perfection of the grown crystals. Remarkably, the refined Flack parameter values for all the measured crystals using a $P4_132$
space-group model were consistently close to either 0 or 1, hence indicating that the considered crystals belong to two enantiomorphic space groups, $P4_132$ and $P4_332$. An anomaly in the resistivity is observed at $\simeq$ 130 K, most likely associated with the onset of a charge-density-wave phase. The superconducting properties (and in particular the symmetry, the amplitude and the temperature dependence of the superconducting gap) were studied by using -- for the first time in this compound -- point contact Andreev-reflection spectroscopy. The results confirm that Mo$_3$Al$_2$C is a moderately strongly-coupled superconductor with $2\Delta/k_\mathrm{B}T_c \simeq 4$ and unambiguously prove that the order parameter has an $s$-wave symmetry despite the asymmetric spin-orbit coupling arising from the lack of inversion symmetry.
\end{abstract}



\maketitle

\section{Introduction}\label{sec:intro}
Noncentrosymmetric superconductors (NCS) have recently attracted considerable attention. The discovery of a mixing of spin-singlet and spin-triplet states in the heavy fermion NCS CePt$_3$Si (with critical temperature $T_c \simeq 0.75$ K \cite{Bauer2004}) has boosted a research aimed at investigating the effects on the superconducting pairing of the lack of inversion symmetry. The latter is reflected in the electronic structure through the spin-orbit coupling; a strong asymmetric spin-orbit coupling indeed results in a spin-splitting of the energy bands \cite{Mineev2012}. Actually, heavy-fermion compounds also display a strong electronic correlation (not to mention the complications due to the localized magnetic moments arising from the rare-earth $f$ electrons) so that they are not the simplest materials where to study the effects of a lack of inversion symmetry. In the past decade, many other NCS with a strong spin-orbit coupling have been investigated, displaying relatively high critical temperatures ($T_c  > 9$ K) and critical magnetic  fields, based on transition metals (Mo, Pd, Pt...) such as, for example, Y$_2$C$_3$ \cite{Amano2004}, Li$_2$Pd$_x$Pt$_{3-x}$B \cite{Togano2004}, Mo$_3$Al$_2$C \cite{Bauer2010} that do not present the aforementioned complications.

Mo$_3$Al$_2$C, in particular, was identified as an interesting candidate for unconventional superconductivity driven by the lack of inversion symmetry \cite{Bauer2010,Karki2010}. The compound has a $\beta$-Mn type structure, with space group $P4_132$ and a $T_c$ of ca.\ 9\,K. Experimental studies of Mo$_3$Al$_2$C polycrystals have been carried out both in the normal and in the superconducting state.
In the normal state, some anomalies in the resistivity, susceptibility, and nuclear-magnetic-resonance (NMR) measurements were interpreted as being due to the opening of a charge-density-wave gap \cite{Koyama2011,Koyama2013} at about 130 K.
In the superconducting state, early results of NMR and specific-heat measurements were interpreted as arising from a non conventional superconducting coupling, possibly with a nodal gap \cite{Bauer2010}. However, the low-temperature specific heat behaviour was found to be exponential, as expected for an isotropic gap \cite{Karki2010}, and  some of the anomalies at higher temperature were later attributed to minor impurity phases. Penetration-depth measurements confirmed this picture evidencing a temperature-independent penetration depth below 0.5 K \cite{Bonalde2011}. In polycrystals of improved quality,  NMR measurements showed a peak in $1/T_1$  just below the transition and a low-temperature exponential behaviour consistent with a pure $s$-wave order parameter \cite{Koyama2013}. More recently, the absence of time-reversal symmetry breaking in Mo$_3$Al$_2$C was concluded by $\mu SR$ measurements \cite{Bauer2014}.

A direct measure of the gap by means of a spectroscopic technique such as tunnel spectroscopy or Andreev-reflection spectroscopy is however missing. In particular, the necessity to carry out direct gap measurements in single crystals has been evidenced by the aforementioned, strong effect of the impurity phases on the measured quantities.

\section{Experimental details and crystal growth}\label{sec:growth}

For the growth of  Mo$_3$Al$_2$C single crystals, we used the cubic-anvil, high-pressure, and high-temperature technique. The details of experimental set up can be found in our previous publications \cite{growth1,growth2}. The apparatus consist of a 1500-ton press, with a hydraulic-oil system comprising a semi-cylindrical multianvil module (Rockland Research Corp.). A set of steel parts transmit the force through six tungsten carbide pistons to the sample in a quasi isostatic way. This method was successfully applied earlier on to grow various compounds, including superconducting and magnetic intermetallic crystals, diamonds, pyrochlores, $Ln$Fe$Pn$O ($ Ln$: lanthanide, $Pn$: pnictogen), oxypnictides, and numerous other compounds \cite{highpressure}.

The main challenges when growing single crystals of Mo$_3$Al$_2$C are the relatively poor reactivity of C and the huge differences in the melting temperatures of components: 2617$^\circ$C for Mo, 660$^\circ$C for Al, and 3827$^\circ$C for C. This is probably the reasons why it is so difficult to synthesize even single-phase polycrystalline samples. As already shown in previous studies, the superconductivity in this material is very sensitive to the details of the heat treatment and to final stoichiometry \cite{Reith2015}. In order to shed light on these metallurgical problems one needs to perform systematic studies on single crystals. We note here that any high-temperature growth in quartz ampoules that utilizes Al as a flux cannot be performed easily, since Al vapor attacks the quartz, leading to a loss of the protective atmosphere. However, as we will show in the present study, the single crystals of Mo$_3$Al$_2$C can be grown successfully by using a high-pressure closed system.

\begin{figure*}[ht]
\includegraphics[keepaspectratio,width=\textwidth]{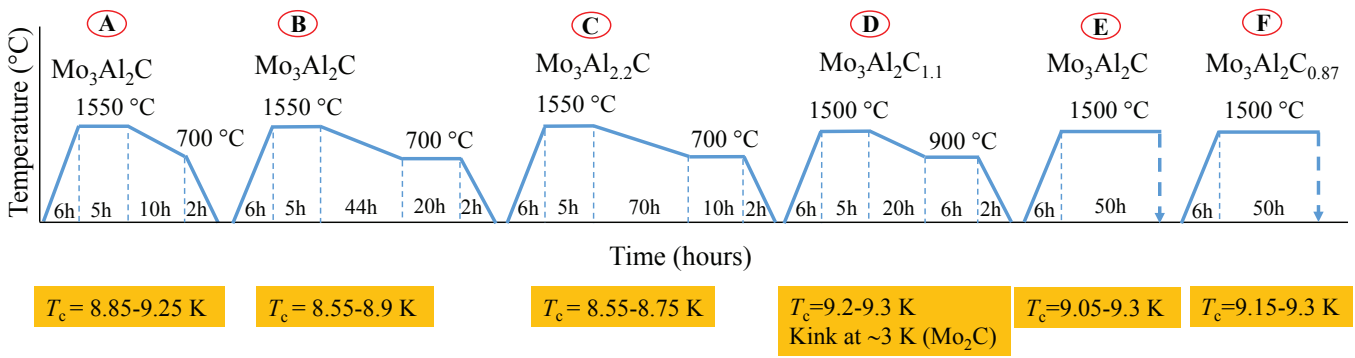}
\caption{The furnace heat-treatment protocols used for the high-pressure and high-temperature synthesis of Mo$_3$Al$_2$C single crystals. The range of critical temperatures ($T_c$) observed for each growth batch is reported in the bottom part of the figure.}\label{fig:protocol}
\vspace{5mm}
\end{figure*}

In order to better understand  the origin of variation in $T_c$ we synthesized more than twenty different compositions with different stoichiometries under high-pressure conditions. The most representative stoichiometries and synthesis protocols are depicted schematically in Fig. \ref{fig:protocol}.

A mixture of Mo, Al, and C powders in a molar ratio Mo$_3$Al$_{2.0-2.2}$C$_{0.87-1.1}$ was placed inside a boron nitride (BN) crucible with an inner diameter of 8.0 mm, and a length of 9.0 mm. The heating element is a graphite tube. Six tungsten carbide anvils generate pressure on the whole assembly. In a typical run, a pressure of 3 GPa is applied at room temperature. The mixture was heat-treated according to the synthesis protocols summarized in Fig.~\ref{fig:protocol} while keeping the pressure constant throughout the growth process. The pressure was released only at the end of the crystal growth. The results of the different loadings and protocols can be summarized as follows:
\begin{itemize}
\item The perfectly stoichiometric Mo$_3$Al$_2$C loadings, heat-treated at 1550$^{\circ}$C and then slowly cooled, resulted in single crystals with $T_c$ ranging between 8.55 and 9.25 K (see Fig. \ref{fig:protocol}, protocols A and B).
\item The Al-rich, i.e., Mo$_3$Al$_{2.2}$C loadings (protocol C) yielded samples with $T_c$ = 8.55 - 8.75 K.
\item The C-rich Mo$_3$Al$_2$C$_{1.1}$ loadings (protocol D) yielded crystals with $T_c$ = 9.2 - 9.3 K, that however contained also a Mo$_2$C phase as impurity besides the Mo$_3$Al$_2$C phase, as revealed through a kink in the magnetic susceptibility at 3 K.
\item The stoichiometric  Mo$_3$Al$_2$C and the C deficient Mo$_3$Al$_2$C$_{0.87}$ loadings heat-treated by protocols E and F led to single-phase samples with $T_c = 9.05 - 9.3$ K and $T_c= 9.15 - 9.3$ K, respectively.
\end{itemize}

The origin of the difference in $T_c$ between different crystals is not clear yet, but current studies suggest that it is most likely due to slightly different sample stoichiometries. Indeed, nominally carbon-deficient crystals (as, e.g., Mo$_3$Al$_2$C$_{0.87}$) feature a higher $T_c$ with respect to nominally stoichiometric ones. This raises the question whether the formation of C vacancies is thermodynamically possible. Considering the well-established crystal structure such an idea looks surprising, as no Mo or Al vacancies could be detected. However, the situation is different for carbon which, as a light atom, makes it difficult to establish experimentally the actual C content. One should also consider that carbides normally exhibit vacancies in their sublattices \cite{Cvacancies}. We note here that polycrystalline Mo$_3$Al$_2$C samples show a rather wide spread of $T_c =9.0 - 9.3$ K, indicating that such samples are not perfectly stoichiometric. All together, these results indicate that the studied compound might be stabilized by vacancy formation.

The influence of stoichiometry, and in particular of carbon deficiency, on the thermodynamic stability and on the electronic structure of Mo$_3$Al$_2$C has been studied theoretically by means of DFT calculations \cite{Reith2015}. The electronic density of states at the Fermi level, which is an important parameter that controls the critical temperature, is enhanced by carbon vacancies. The phonon DOS $D(\omega)$ is heavily affected as well: on increasing the concentration of carbon vacancies, the low-energy part of the spectrum becomes increasingly Debye-like [$D(\omega) \propto \omega^2$]. Although a detailed evaluation of the effect of such changes on the critical temperature requires a precise knowledge of the electron-phonon spectral function $\alpha^2F(\omega)$ (rather than of the phonon DOS alone), the link between the concentration of C vacancies and the critical temperature is highly plausible.

In our experimental conditions we observed variations in $T_c$ even within individual batches. Therefore, besides the starting nominal composition and synthesis temperature, the temperature gradient across the sample is another important parameter that controls $T_c$. For a cell temperature of $1500-1550^{\circ}$C, we estimate a temperature gradient across the sample of ca.\ $70^{\circ}$C. The whole assembly produces, therefore, a parabolic temperature profile across the sample, with the maximum temperature being reached at its center (for more details, see Ref. \cite{growth1}). Thus, one may expect the formation of Mo$_3$Al$_2$C with slightly lower $T_c$s at the top and bottom parts of the BN crucible, whereas crystals with a higher $T_c$, i.e., Mo$_3$Al$_2$C$_{0.87}$, will grow close to the centre of the crucible, where the temperature is highest. Our experimental findings are consistent with DFT calculations reported in Ref. \cite{Reith2015} and demonstrate that superconducting transition temperature and carbon vacancy concentration can be controlled by the synthesis protocol. It is quite plausible that quenching of the samples from high temperatures will result in the freezing of high concentration of C vacancies ($\simeq$ 0.13 - 0.14), whereas at slow cooling process the vacancy concentration might be reduced since more carbon atoms can be introduced in the lattice.

\begin{figure}
\includegraphics[width=\columnwidth]{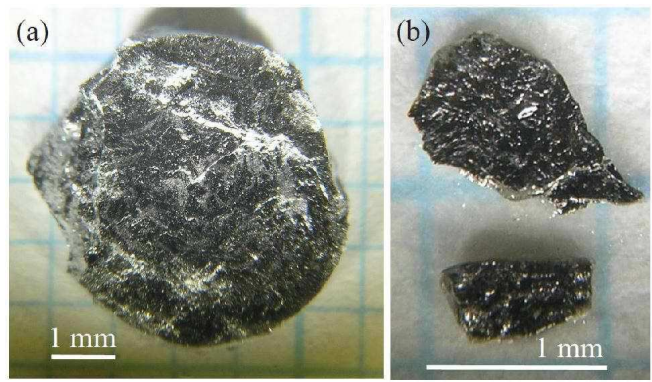}
\caption{Optical microscope images of Mo$_3$Al$_2$C crystals. (a) Top view of as-grown solidified lump after radial crushing. The lump was limited by the crucible and the circular cross section is due to the crucible walls. (b) Mechanically extracted Mo$_3$Al$_2$C crystals.}
\label{fig:crystals}
\end{figure}

Figure \ref{fig:crystals} shows optical microscope images of  Mo$_3$Al$_2$C single crystals. Fig. \ref{fig:crystals}a shows a top view of the as-grown solidified lump with a mixture of single-crystalline Mo$_3$Al$_2$C and some flux. The lump was crucible-limited and its circular cross section is due to the crucible walls. After crushing, a large number of crystals with various shapes, and sizes up to $1\,\mathrm{mm}^3$ were found (Fig. \ref{fig:crystals}b).
The stoichiometry of the grown crystals was checked by energy-dispersive X-ray spectroscopy. For each growth batch, 5 to 7 crystals were analyzed. Their Mo and Al content was always found to be close to the stoichiometric 3:2 composition within experimental resolution 0.1. The search for possible C deficiencies in nominally deficient samples did not prove conclusive due to substantial fluctuations of the experimental points between 0.8 and 1.0.

\begin{figure}
\includegraphics[width=\columnwidth]{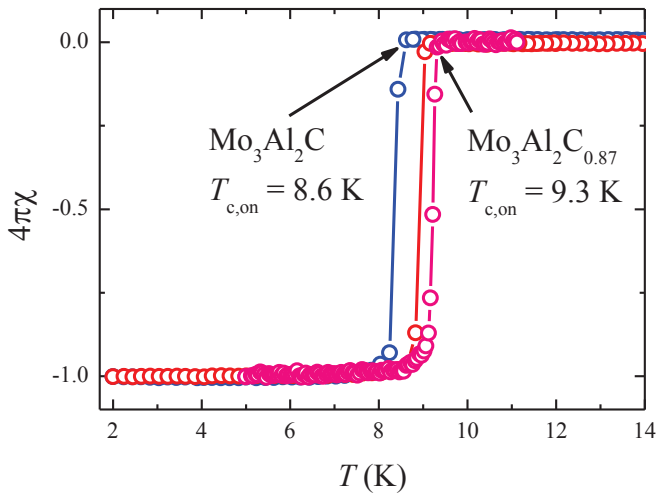}
\caption{Zero-field-cooled magnetization curves for Mo$_3$Al$_2$C crystals extracted from various growth batches. The superconducting transition is rather narrow and the critical temperature varies between 8.6 and 9.3 K.}
\label{fig:magnetization}
\end{figure}

A magnetic determination of the critical temperature for each growth batch was performed on clean single crystals with sizes of few hundreds of micrometers. Some examples of zero-field-cooled (ZFC) temperature-dependent measurements in low field are shown in Fig. \ref{fig:magnetization}. The spread of $T_c$ for each individual batch is indicated in Fig. \ref{fig:protocol}. In general, the transition to the superconducting state is very narrow; the ``10-90'' criterion gives $\delta T_c = T_c^{90}-T_c^{10} \simeq 0.3$ K. The good quality of the grown crystals is confirmed by the resistivity studies presented in section IV.

\section{Structure refinement}\label{sec:structure}
Single-crystal X-ray diffraction experiments were performed with an Oxford Diffraction diffractometer equipped with an Onyx CCD detector, a Mo X-ray tube and a graphite monochromator adjusted to select K lines. Several crystals were measured to obtain  representative crystal structures. The measurements at $\lambda=0.71073$ {\AA} were performed with an oscillation angle of 1$^{\circ}$. Reflection integration and data reduction were performed using the software package CrysAlis from Oxford Diffraction. Structure solution and the determination of atomic positions were performed using the programs SUPERFLIP \cite{Palatinus2007} and EDMA \cite{Palatinus2012} respectively. All the refinements were done with SHELXL97 \cite{Sheldrick2008}.

\begin{figure}
\includegraphics[width=\columnwidth]{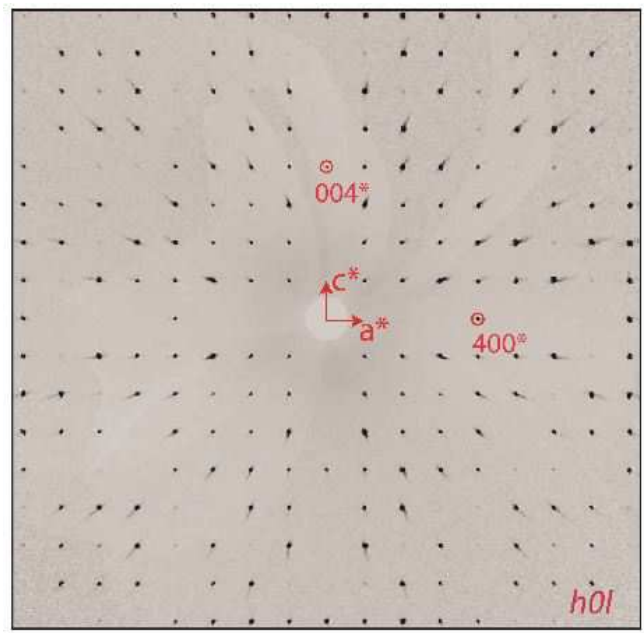}
\caption{Reconstructed $h0l$ layer of the reciprocal space. Red circles mark some of reflections that satisfy the reflection condition $h00: h=4n$.}\label{fig:Xray}
\end{figure}

\begin{figure}
\vspace{3mm}
\includegraphics[angle=-90,keepaspectratio,width=0.95\columnwidth]{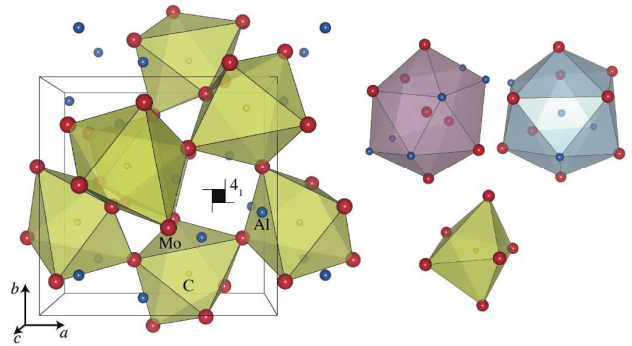}
\caption{Crystal structure of Mo$_3$Al$_2$C (space group: $P4_132$) as viewed along the $c$-axis (left) and coordination polyhedra of the constituent atoms (right).}\label{fig:structure}
\end{figure}

The Fourier spectra of all the measured crystals exhibit an $m\overline{3}m$ Laue symmetry. The reflection conditions we find ($h00$: $h = 4n$) correspond to two enantiomorphic space groups, $P4_132$ and $P4_332$ (Fig.\ref{fig:Xray}). These can be distinguished based on the Flack parameter value \cite{Flack1983}, provided the anomalous signal is strong enough, a sufficient number of Friedel pairs is measured, and special attention is paid to the absorption corrections. These factors are critical to achieve a low estimated standard deviation (\emph{e.s.d.}) for the Flack parameter and, hence, a reliable conclusion about the absolute crystal structure. To facilitate the accurate refinement of the Flack parameter, datasets with a high redundancy and a coverage close to 100 were collected for all the crystals. Regardless of all the precautions adopted, the minimal achieved \emph{e.s.d.}'s for the Flack parameter value were about 0.1, which is somewhat higher than the recommended value of less than 0.04. However it is remarkable that Flack parameter values refined for all the measured crystals using the model with the space group $P4_132$ were reproducibly close to either 0 or 1, a strong indication that the studied samples were a mixture of crystals with space groups $P4_132$ and $P4_332$. The results of two representative refinements are reported in Table \ref{tab:1} and in Ref. \cite{suppl}. The refined atomic positions and displacement parameters for the model with space group $P4_132$ are presented in Tables \ref{tab:2} and \ref{tab:3}.

\begin{table*}
\begin{tabular}{|l|l|l|}
\hline
\multicolumn{3}{|c|}{Crystal data} \\
\hline
Chemical formula    &	Mo$_3$Al$_2$C   &	Mo$_3$Al$_2$C \\
\hline
$T_c$ (prot. B)  &   8.9 K    &   8.9 K  \\
\hline
$M_r$ (g/mol)	    &     353.79	     &  353.79 \\
\hline
$Z$ & 	4	&  4 \\
\hline
Cell setting, space group &	Cubic, $P4_132$	& Cubic, $P4_332$ \\
\hline
$a$ ({\AA}) 	& 6.86715(3)	& 6.86793(3) \\
\hline
$V$ ({\AA}$^3$)	& 323.839(3)	&  323.950(3)\\
\hline
$\mu$ (mm$^{-1}$)	&  11.697	&   11.697 \\
\hline
Crystal shape, color &	Platelet, black & 	Platelet, black \\
\hline
Crystal size (mm$^3$)	&  $0.08 \times 0.06 \times 0.02$	&  $0.06 \times  0.04 \times 0.02$ \\
\hline
Temperature (K) & 	298 	&  298 \\
\hline
Radiation type & 	Mo$K \alpha$ &	Mo$K \alpha$ \\
\hline
Radiation wavelength ({\AA})	& 0.71073	& 0.71073 \\
\hline
\multicolumn{3}{|c|}{Data collection} \\
\hline
Diffractometer	& Goniometer Xcalibur, & Goniometer Xcalibur \\
\hline
Detector &  Onyx CCD & Onyx CCD \\
\hline
Data collection method &	  scans &	  scans \\
\hline
Scan width (deg)	& 1.0	&  1.0 \\
\hline
Absorption correction &	Multi-scan &	Multi-scan \\
\hline
$T_{min}$ &	0.42981 &	0.72247 \\
\hline
$T_{max}$	&  1.0000 &	1.0000 \\
\hline
Number of measured reflections & 11189 & 12086 \\
\hline
Number of independent reflections &   293 & 294 \\
\hline
Number of observed reflections &  289 &  289 \\
\hline
Criterion for observed reflections &  $ I > 2 \sigma (I)$ & $ I > 2 \sigma (I)$ \\
\hline
$R_{int}$(all), $R \sigma$(all) & 0.0483, 0.090 & 0.0339, 0.0070 \\
\hline
$\theta_{max}$ (deg) & 	37.6	&  37.6 \\
\hline
\multicolumn{3}{|c|}{Refinement} \\
\hline
Refinement on &	$F^2$	&  $F^2$ \\
\hline
Weighting scheme &	$w = 1/[\sigma^2(F_o^2)+(0.0178 P)^2+0.8687 P]$, &	$w = 1/[\sigma^2(F_o^2)+(0.0233 P)^2+0.2695 P]$ \\
 & where $P=(F_o^2+2F_c^2)/3$ & where $P=(F_o^2+2F_c^2)/3$ \\
\hline
$R_{obs}$ / $R_{all}$ / $wR_{obs}$ / $wR_{all}$ &	0.0170/0.0175/0.0378/0.0379 & 	 0.0157/0.0167/0.0370/0.0373 \\
\hline
Goodness-of-fit (obs/all)	&  1.196/1.196 &	1.208/1.208 \\
\hline
No. of parameters	&  13	&  13 \\
\hline
$(\Delta/\sigma)_{max}$ & 	0.825 &	0.176 \\
\hline
$\Delta \rho_{max}$, $\Delta \rho_{min}$ (e {\AA}$^{-3}$) &	0.681, -0.778	&  0.842, -0.931 \\
\hline
Extinction correction	&	SHELXL97	&  SHELXL97 \\
\hline
Extinction coefficient &	0.0276(18) &	0.0122(14) \\
\hline
Absolute structure	& Flack(1983) &	Flack(1983) \\
\hline
Flack parameter &	-0.07(13) &	-0.02(11) \\
\hline
\end{tabular}
\caption{Experimental and refinement details. Computer programs used: CrysAlis CCD (Oxford Diffraction, 2008), SHELXL97 (Sheldrick, 2008).} \label{tab:1}
\end{table*}

\begin{table*}
\begin{tabular}{|l|l|l|l|l|l|}
\hline
Atom	&  Wyckoff position	 &  $x$	&  $y$	& $z$	&  $U_{iso}/U_{eq}$ ({\AA}$^2$) \\
\hline
Mo	&  12$d$	& $1/8$ &	0.79721(3) &	0.04721(3) &	0.00933(10) \\
\hline
Al	& 8$c$ &	0.93283(10)	&  0.43283(10) &	0.06717(10) &	0.0067(2)\\
\hline
C	& 4$b$ &  $3/8$ &	$5/8$ & $1/8$ &	0.0104(8) \\
\hline
\end{tabular}
\caption{Wyckoff positions, atomic coordinates and isotropic/equivalent isotropic displacement parameters for the model with space group $P4_132$.}\label{tab:2}
\end{table*}

\begin{table*}
\begin{tabular}{|l|l|l|l|l|l|l|}
\hline
	&  $U^{11}$     &	$U^{22}$   &	$U^{33}$   &	$U^{23}$   &	$U^{13}$	  & $U^{12}$ \\
\hline
Mo	& 0.01304(14)	& 0.00748(10)	& 0.00748(10) &	-0.00031(8)	& -0.00190(6)	& 0.00190(6)\\
\hline
Al	& 0.0067(2) &	0.0067(2) &	0.0067(2) &	-0.00037(19)	& -0.00037(19)	&  0.00037(19) \\
\hline
C	& 0.0104(8)	 & 0.0104(8) &	0.0104(8)	& -0.0014(9) &	0.0014(9)	& 0.0014(9)\\
\hline
\end{tabular}
\caption{Atomic displacement parameters  ({\AA}$^2$)} \label{tab:3}
\end{table*}

The crystalline structure of Mo$_3$Al$_2$C corresponds to that of $\beta$-Mn: both Mo and Al atoms are in a distorted icosahedron,  while C atoms are in a distorted octahedral coordinations built from metal atoms (Fig.~\ref{fig:structure}). The space groups $P4_132$ and $P4_332$ differ by inversion symmetry, arising from the pair of enantiomorphic axes $4_1$ and $4_3$. The difference between the two axes can be explained with reference to Fig.~\ref{fig:structure}. By considering the $z$-coordinate of the Al-atom closest to $4_1z$-axis, its symmetry-equivalent is obtained via a counterclockwise rotation by $90^{\circ}$ with subsequent translation by $c/4$ along the $c$-axis. The translation component of the $4_3$ $z$-axis is the same, but the rotation occurs clockwise. Hence, in the former case Al atoms (as well as Mo and C) ascend counterclockwise along the $c$-axis, whereas in the latter case Al atoms ascend clockwise along the $c$-axis. The same is valid for the coordination polyhedra. To further confirm the enantiomorphic purity of the studied crystals, single crystal X-ray diffraction at the other wavelengths or at lower temperature could be performed.

\section{Electric transport measurements}
\begin{figure}
\includegraphics[width=\columnwidth]{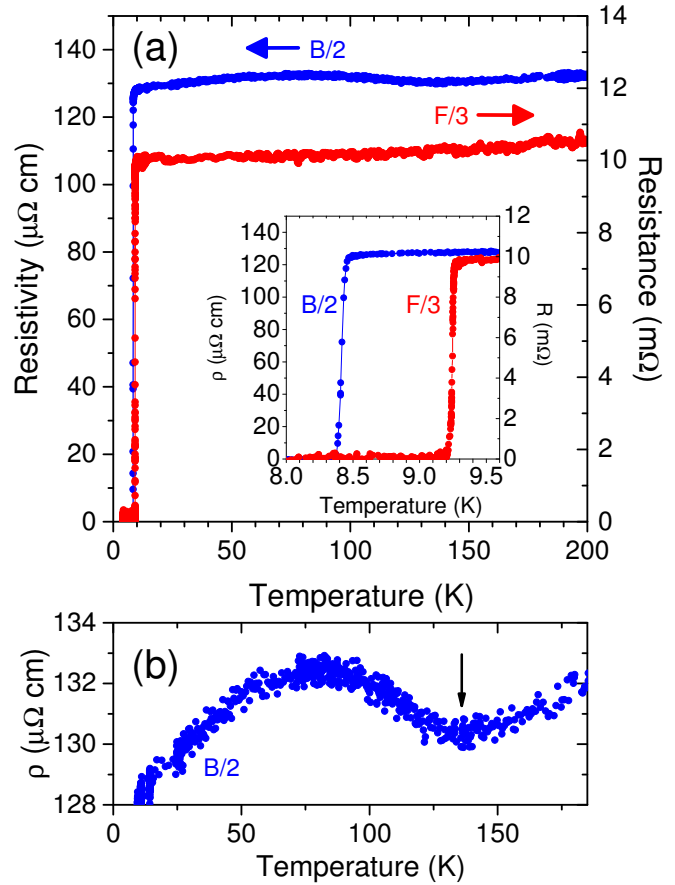}
\caption{(a) Resistivity of the crystal B/2 (blue circles, left-hand vertical axis) and resistance of the crystal F/3 (red circles, right vertical axis) as a function of temperature up to 200 K. Inset: detail of the superconducting transitions. (b) Magnification of the normal-state resistivity of B/2 showing a clear minimum around 130 K.}\label{fig:res}
\end{figure}

Transport measurements were carried out as a function of temperature either in vacuum (using a Cryomech$^\mathrm{TM}$ ST-403 3K pulse-tube cryocooler) or in He exchange gas, by mounting the crystals on the cold head of a cryogenic insert.
Figure \ref{fig:res}a shows the resistivity of crystal B/2 (i.e., obtained with the protocol B, see Fig.~\ref{fig:protocol}) as a function of temperature (blue circles). The residual resistivity is $\rho_0 = 125 \, \mu \Omega$ cm, i.e., of the same order of magnitude as that reported for some polycrystalline samples \cite{Bauer2010} but, unlike in polycrystals \cite{Bauer2010,Karki2010,Koyama2013}, a rather clear minimum in the resistivity is observed at $T^*=134 \pm 4$ K (see Fig.~\ref{fig:res}b). It is interesting to note that the opening of a charge density wave (CDW) gap at about 130 K was proposed \cite{Koyama2011} based on $^{27}\mathrm{Al}$ nuclear magnetic resonance measurements that showed anomalies in the Knight shift and in the spin-lattice relaxation rate \cite{Kuo2012}.  A signature of these anomalies was later found in the resistivity of polycrystals in the form of a slope change in the $\rho(T)$ curve \cite{Koyama2013}.

As clearly shown in Fig.~\ref{fig:res}a, the superconducting transition of the B/2 crystal is rather sharp, with an onset at $T_{c}^{on}=8.48 \pm 0.02$ K and a zero-resistance state achieved at $T_c^0=8.35\pm 0.02$ K. Note that the critical temperature agrees very well with that determined via magnetization ($T_c^{M,on}=8.55$ K) and, as in all the crystals grown with protocol B, is smaller than that reported in the literature for this compound ($T_c=9.3$ K).

Fig.~\ref{fig:res}a reports also the resistance of the C-deficient crystal F/3 (grown by using the protocol F) as red symbols; the irregular shape of the crystal did not allow us to reliably estimate the resistivity. In this case $T_c^{on}= 9.27 \pm 0.02$ K and $T_c^0=9.19 \pm 0.02$ K. The critical temperature resulting from magnetization measurements in this case was $T_c^{M,on} = 9.3$ K,  again in perfect agreement with the transport results. The $R(T)$ curve in this case is monotonic, i.e., does not exhibit any minimum around 130 K. At this temperature, a smooth slope change is instead observed, as in polycrystals \cite{Koyama2013}.

\begin{figure}
\includegraphics[width=0.9\columnwidth]{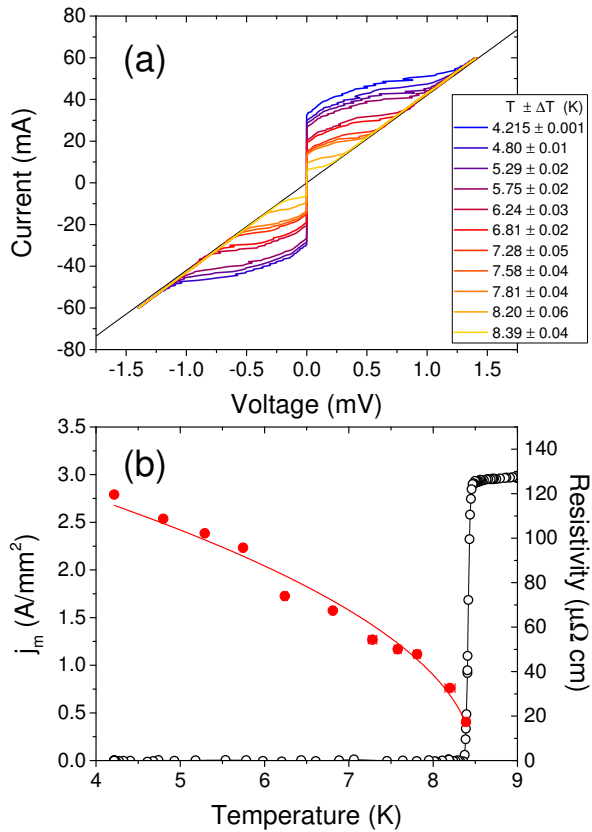}
\caption{(a) I-V characteristics of the B/2 crystal, as a function of temperature. (b) Red solid circles: values of the pseudocritical current $j_m$ (minimum normal-zone existence current density) that marks the onset of an electric field of $1 \mu$V/cm (left scale). The solid line is a fit of the data with a function of the form $j_m=C \sqrt{1-(T/T_c)}$ that gives $T_c=8.48 \pm 0.05$ K. Black open circles: resistivity of the same crystal as a function of temperature (right scale).  Note that $T_c^{on}= 8.48 \pm 0.02$ K. }\label{fig:critcurr}
\end{figure}

Figure~\ref{fig:critcurr}a shows the temperature dependence of the $I$-$V$ characteristics of the  B/2 crystal. The measurements were performed by applying short current pulses and measuring the voltage drop across the sample in the same four-probe configuration used for the resistivity measurements. To optimize the thermal exchange between the sample and its environment, allow a proper heat dissipation, and ensure that every single point was taken at the same starting temperature conditions, the sample was mounted in He exchange gas and the current pulses were separated by several seconds of idle time.
The curves clearly show the current-induced transition from the superconducting to the resistive state, and then to the normal state, where their slopes corresponds to the normal-state conductance of the sample just above $T_c$ (straight line). Figure~\ref{fig:critcurr}b shows the temperature dependence of the current density (let us call it $j_m$ for reasons that will become clear below), at which the electric field across the sample is 1\,$\mu$V/cm. In other words, $j_m$ is the current density at which a non-zero resistance appears. Its temperature dependence is \cite{gurevich1987}:
\begin{equation}
j_m(T) = C \sqrt{1-\frac{T}{T_c}}.
\end{equation}
The fit of the data points with this function (solid line in Fig.~\ref{fig:critcurr}b) gives $T_c = 8.48$ K, corresponding to $T_c^{on}$, i.e., to the achievement of the normal state in the resistivity curve (see Fig.~\ref{fig:res}a). This functional form of the temperature dependence of $j_m$, together with the shape of the I-V curves themselves (the very sharp onset of a finite voltage, the ubiquitous ripples in the transition region, the linear behaviour just above $j_m$) clearly indicate that the transition to the resistive state is dominated by self-heating effects and thus occurs at current densities $j_m$ much smaller than the real $j_c$ \cite{gurevich1987}. As we will show in the following, this is relevant for the analysis of the results of point-contact Andreev-reflection spectroscopy.

\section{Point-contact Andreev reflection spectroscopy (PCARS)}
Point-contact spectroscopy is a simple but very effective technique for the measurement of the amplitude and symmetry of the energy gap in superconductors. It consists in measuring the differential conductance ($dI/dV$) of a ballistic contact between a normal metal (N) and the superconductor (S) under study, as a function of the voltage bias. If the potential barrier at the N/S interface  is sufficiently small, the conduction is dominated by Andreev reflection \cite{Andreev,BTK}, even though the probability of quasiparticle tunnelling is not zero. The spectrum obtained in this way can be fitted to suitable models for the Andreev reflection at the N/S interface \cite{BTK,kashiwaya} to obtain quantitative information on the amplitude of the gap and on its symmetry in the reciprocal space.
The condition for ballistic conduction (and thus for energy-resolved information to be extracted from the spectra) is that the radius of the contact (schematized as a circular aperture in the otherwise completely opaque barrier between the N and S banks) is smaller than the electronic mean free path in both banks, i.e. $a \ll \mathrm{min}[\ell, \ell']$. When this condition is satisfied, the resistance of the contact (called Sharvin resistance) is determined only by the geometry, i.e., by the number of the quantum-conductance channels:
\begin{equation}
R_N = R_S = \frac{2h}{e^2 a^2 k_{F, min} \tau} \label{eq:Sharvin}
\end{equation}
where $\tau$ is a function of the Fermi velocities in the two banks \cite{daghero2010} and is often disregarded for approximate estimations. Note that $R_N$ is the resistance of the ballistic contact when both banks are in the normal state, but also coincides with the resistance of the same contact in the superconducting state ($T<T_c$) measured at sufficiently high bias (i.e., $V > 3\Delta$, with $\Delta$ the amplitude of the superconducting gap) \cite{daghero2010}.
%

Point contacts on Mo$_3$Al$_2$C crystals were made by using the so-called ``soft'' PCARS technique, i.e., by stretching a thin gold wire across the crystal (that was not subject to any treatment), so that it touches the crystal in a single point. Sometimes we used a drop of Ag paste to make the contact, which ensures a much higher mechanical and thermal stability of the contact itself \cite{daghero2010}; in other cases we used the gold wire alone. In each case, Mo$_3$Al$_2$C represents the bank with the smaller $k_F$.  By using the free-electron approximation one finds that the Sharvin resistance reads:
\begin{equation}
R_S = \frac{4 \rho \ell}{3 \pi a^2}
\end{equation}
where both $\rho$ and $\ell$ refer to Mo$_3$Al$_2$C. The mean free path has been reported to be $\ell = 3$ nm \cite{Bauer2010} and the low-temperature resistivity is $\rho \simeq 125 \mu\Omega$cm, so that the condition $a < \ell$ means that the resistance of the contact should be larger than 180 $\Omega$ to fulfill the ballistic conditions.
Obtaining such a high value of contact resistance is not straightforward because the surface of the crystals is extremely clean.

\begin{figure}
\includegraphics[width=0.9\columnwidth]{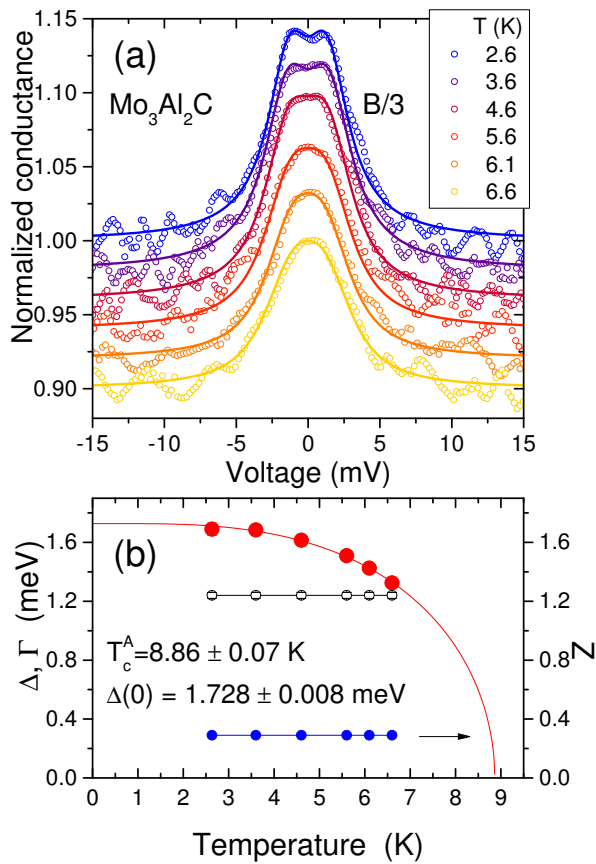}
\caption{(a) Normalized conductance curves of a Au/Mo$_3$Al$_2$C contact (on the crystal B/3) with $R_N =250 \, \Omega$, recorded at different temperatures (symbols) with the relevant fit with a 2D BTK model \cite{kashiwaya} for a $s$-wave order parameter. (b) Temperature dependence of the energy gap extracted from the fit (red circles) and of the other parameters, $Z$ (blue squares) and $\Gamma$ (black open circles). Note that neither $\Gamma$ or $Z$ depend on temperature, as expected for a perfectly ballistic contact.}\label{fig:goodconductance}
\end{figure}

Figure \ref{fig:goodconductance}a shows some conductance curves, recorded at different temperatures (starting from 2.6 K, i.e., below $T_c/3$) of a Au/Mo$_3$Al$_2$C contact with $R_N = 250\, \Omega$ in the crystal B/3 that has a critical temperature $T_c^{M,on} = 8.75$ K determined from magnetization. The enhancement of conductance at low bias is due to Andreev reflection; the shape of the curves is typical of an $s$-wave superconductor, without any evidence of other components of the order parameter such as, for instance, a $p$-wave one. This is fully consistent with previous findings from magnetic penetration depth measurements \cite{Bonalde2011} and muon spin rotation and relaxation studies \cite{Bauer2014}, that showed the absence of time-reversal symmetry breaking in Mo$_3$Al$_2$C. The two symmetric maxima occur roughly at $V= \pm \Delta/e$, but a fit with a suitable model was required to gather more accurate quantitative information. To this aim, the curves were normalized, i.e., divided by $R_N$, and vertically shifted for clarity; the solid lines in the figure represent best fits using the Tanaka-Kashiwaya version \cite{kashiwaya} of the BTK model \cite{BTK} for Andreev reflection at a N-S interface. The agreement, especially in the low-bias region, where the noise is smaller, is very good. The model we used contains only three parameters: the gap amplitude $\Delta$, a barrier parameter $Z$, and a smearing parameter $\Gamma$ that accounts for the intrinsic finite lifetime of quasiparticles and for extrinsic (experimental) sources of broadening \cite{daghero2010}. Note that only $\Delta$  was varied as a function of temperature to adjust the theoretical curves to the experimental spectra, while both $Z$ and $\Gamma$ were determined from the fit of the lowest-temperature curve and then kept constant on increasing the temperature.

The temperature dependence of the gap extracted from the curves is shown in Fig.~\ref{fig:goodconductance}b. Unfortunately, the contact broke down (because of the different thermal expansion coefficients of the various materials) at 6.6 K. However, the data points could still be fitted to a BCS-like curve:
\[\Delta(T) = \Delta(0) \tanh \left(1.74 \sqrt{\frac{T_c}{T}-1}\right)\]
giving $\Delta(0) = 1.728 \pm 0.008$ meV and $T_c=8.86 \pm 0.07$. The corresponding gap ratio $2 \Delta(0) /k_B T_c = 4.53$ lies midway between those obtained from specific-heat measurements (4.028 \cite{Karki2010} and 4.06 \cite{Bauer2014}) and that given by penetration depth measurements (5.18 \cite{Bauer2014}).

\begin{figure}
\includegraphics[height=\columnwidth,angle=-90]{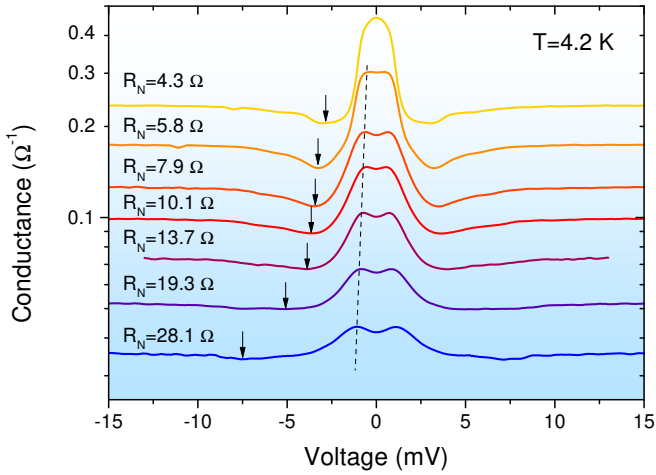}
\caption{Differential conductance curves of a Mo$_3$Al$_2$C / Ag-paste contact for different values of the contact resistance $R_N$, that was progressively reduced by applying short voltage pulses, as explained in Ref.\cite{daghero2010}. Note the logarithmic scale on the vertical axis. Arrows indicate the position of the minima in the conductance (i.e. the dips).}\label{fig:conductances}
\end{figure}

In contacts with resistance smaller than 180 $\Omega$, the spectra systematically display anomalous ``dips" at intermediate bias values, whose position strongly depends on the resistance of the contact. As an example, Fig.~\ref{fig:conductances} reports the conductance curves of the \emph{same} Mo$_3$Al$_2$C/Ag paste contact on the crystal F/3 ($T_c^{M,on}=9.3$ K determined from magnetization) for different values of its resistance, progressively reduced by applying short voltage pulses as described elsewhere \cite{daghero2010}. It is clear that, on decreasing $R_N$, the dips become deeper and deeper and move towards lower voltages. This behavior excludes that such dips have any relation with the superconducting energy gap or with any other intrinsic energy scale of the material under study.

There are two possible mechanisms that can give rise to these dips.
The first, in principle, can occur even for ballistic contacts, and reflects simply the fact that, on increasing the current, the critical current density (or, better, $j_m$, as shown in Fig.~\ref{fig:critcurr}) is reached and the material turns to the normal state. This picture, nevertheless, does not apply to our case since, in purely ballistic contacts (see Fig.\ref{fig:goodconductance}), such dips are missing.

The other mechanism for the formation of the dips relates them to the breakdown of the  conditions of ballistic conduction through the contact \cite{Sheet2004,Daghero2006}. A rough estimate of the contact radius for the curves of Fig.~\ref{fig:conductances}, based on the Sharvin equation, indicates that $a$ increases from $\simeq 7.5$ nm when $R_N = 28.1\, \Omega$ to $\simeq 19$ nm when $R_N = 4.3 \, \Omega$. Since $\ell \simeq 3$ nm, none of these contacts is actually in the Sharvin regime. The contact resistance can thus be approximately expressed \cite{Wexler1966} as $R_N = R_S + R_M$ where the Maxwell term $R_M$ is equal to $(\rho + \rho')/2a$. In the superconducting state at low $T$, $\rho'=0$ and the Maxwell term only contains the (very small) resistivity of gold. Since $R_S \propto a^{-2}$ and $R_M \propto a^{-1}$, their relative contribution to the total contact resistance depends on $a$. Unless $a$ is very small, $R_M$ is not negligible and this has many consequences.

\begin{figure}
\includegraphics[width=0.9\columnwidth]{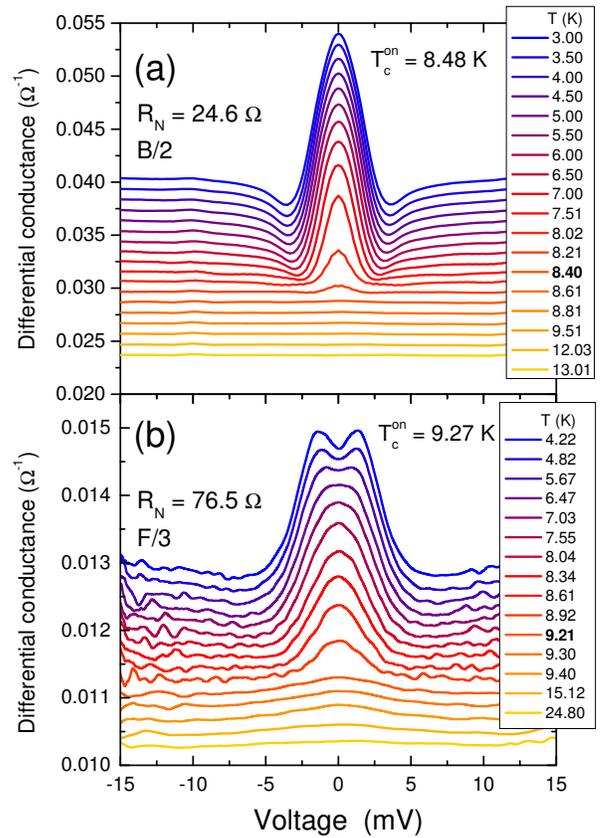}
\caption{Two examples of temperature dependencies of PCARS spectra.  (a) Au/Mo$_3$Al$_2$C contact on crystal B/2, having $R_N=24.6 \, \Omega$ and a large Maxwell contribution so that the Andreev features are almost completely eroded by the dips, and the zero-bias feature is unnaturally enhanced with respect to the high-bias tails. (b) Ag-paste Mo$_3$Al$_2$C contact on crystal F/3, having $R_N=76.5 \, \Omega$, and a smaller Maxwell contribution to the resistance. The dips here fall sufficiently far from the edges of the Andreev-reflection feature, at least at low temperature. In both (a) and (b) all the curves apart from the top one are vertically offset for clarity. The temperature corresponding to the achievement of the normal state is indicated in bold.} \label{fig:Tdep}
\end{figure}

First, when the critical conditions of temperature and current density are reached,  $\rho'$ starts playing a role in the Maxwell term. Since $\rho'$ is again proportional to the derivative of the curves in Fig.~\ref{fig:critcurr}, the dips arise \cite{Sheet2004}. Clearly, the smaller the contact resistance, the smaller the voltage at which this happens, as shown in Fig.~\ref{fig:conductances}. When the dips are too close to $\Delta/e$, the Andreev-reflection features look narrower than in higher-resistance contacts; in the limit case, they collapse into a single featureless maximum at zero bias (top curve in Fig.~\ref{fig:conductances}). Moreover, the dips move towards lower voltages in curves recorded at increasing temperatures. This is shown for example in Fig.~\ref{fig:Tdep}a that reports the temperature dependence of the conductance curves of a point contact on the crystals B/2, with a resistance $R_N \simeq 24.6\, \Omega$.

Second, once $\rho' \neq 0$, Joule heating in the contact starts to be substantial and rapidly makes the temperature in the contact increase according to the law $T^2= T_0^2 + V^2/4L$, $L$ being the Lorentz number and $T_0$ the bath temperature. Therefore, only the part of the spectrum at $|V|<V_{dip}$ is actually recorded at the bath temperature. Third, while the resistance of the contact at low bias is practically $R_S$ (the term $\rho/4a$ being very small) it becomes equal to $R_S+R_M$ at high bias. Since $R_M/R_S$ increases on increasing $a$, in low-resistance contacts the low-bias part appears enhanced with respect to the tails and when such curves are normalized their amplitude can sometimes exceed the ideal value of 2. The two top curves in Fig.\ref{fig:conductances} are close to this condition.

The conclusion of this digression is that a reliable determination of the gap amplitudes requires ballistic contacts, but even the spectra of diffusive contacts can provide a reasonably good estimation of the gap amplitude, provided that the dips fall sufficiently far from $\Delta/e$. For example, the curves shown in Fig.~\ref{fig:Tdep}b are at the limit. At low temperature the dip falls at a voltage high enough to ensure that the central part of the curve (i.e. up to and slightly above the conductance maxima) is still reliable, and its fit can provide a reasonable estimation of the gap.

\begin{figure}
\includegraphics[width=0.9\columnwidth]{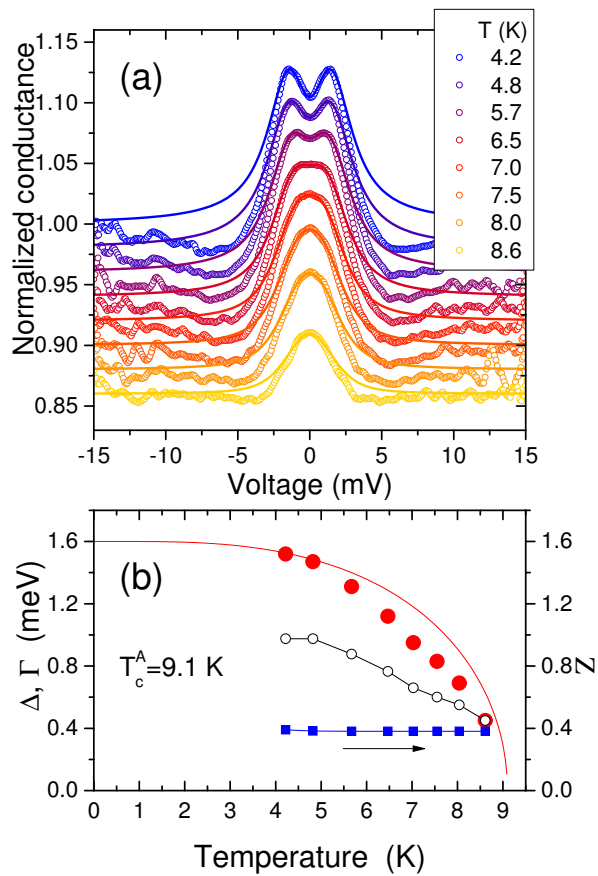}
\caption{(a) The PCARS spectra obtained by normalizing the differential conductance curves of Fig.\ref{fig:Tdep}b (symbols) together with the relevant fit  within the 2D BTK model. The best fit was obtained by minimizing the SSR in range $\pm 2$ mV. (b) Temperature dependence of the best-fitting parameters $\Delta$ (solid circles) $\Gamma$ (open circles) and $Z$ (solid squares).} \label{fig:Tdepfit}
\end{figure}

The fit of the curves reported in Fig.~\ref{fig:Tdep}b is shown in Fig.~\ref{fig:Tdepfit}. In panel (a) the PCARS spectra (i.e. the differential conductance curves normalized to their high-voltage resistance) are compared to the best-fitting 2D-BTK curve. The fitting procedure was the following. First we approximately determined up to which voltage the model for an ideal contact was able to account for the observed shape of the curve. In other words, we excluded from the calculation of the sum of squared residuals (SSR) the part of the curve dominated by the dips. It turned out that a safe range was $\pm 2$ meV. This is enough to fit the region of the zero-bias valley and the two symmetric maxima -- which is the most important to assess the possible presence of zero-bias excitations associated to gapless regions of the Fermi surface (like, e.g. when a gap has nodes) or more generally the symmetry of the gap in the $k$ space.

Then, we fitted all the curves by minimizing the SSR in the same range at all temperatures. This was made both by using an automatic fitting procedure using the 1D BTK model \cite{BTK}, and by changing the parameter values by hand, within the 2D BTK model \cite{kashiwaya,daghero2010}. In both cases, we used an isotropic $s$-wave gap. Within the range of reliability, there is again no detectable discrepancy between the experimental curves and the theoretical spectra, and hence no sign of a non-$s$-wave component. The values of the gap amplitude $\Delta$ and of the broadening parameter $\Gamma$ resulting from the two procedures were identical ($Z$ values were instead different, as expected, because of the different assumptions of the two models).

Fig.~\ref{fig:Tdepfit}b shows the fitting parameters resulting from the 2D BTK model.
Because of the presence of the dips and of their progressive moving to lower voltages on increasing temperature, the values of the gap start to deviate very soon (at about 6 K) from a BCS-like $\Delta(T)$ curve. Therefore, determining the zero-temperature gap amplitude $\Delta_0$ is not easy: Qualitatively, one can say that only the first two points lie on top of a BCS-like curve with $\Delta_0=1.6$ meV and gap ratio 4.1. Note that the values of $\Gamma$ necessary to obtain a good fit decrease on increasing temperature. This is a clear sign that the Andreev features are progressively narrowing, eroded by the dips. In the ideal case $\Gamma$ should instead increase or at most remain approximately constant (see Fig.~\ref{fig:goodconductance}b).

\begin{figure}
\includegraphics[height=0.9\columnwidth,angle=-90]{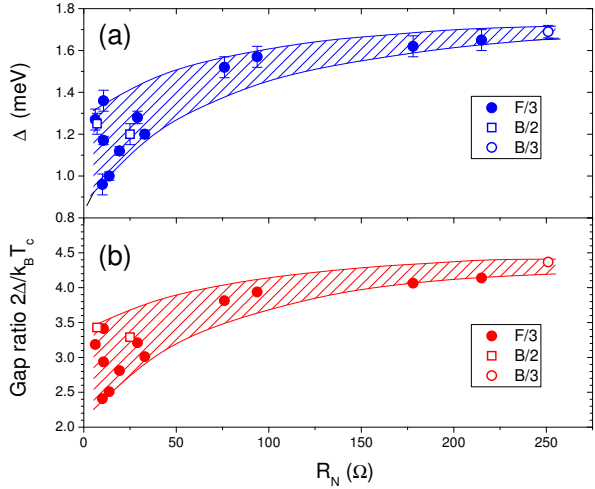}
\caption{(a) Amplitude of the energy gap $\Delta$ extracted from the fit of low-temperature conductance curves, as a function of the contact resistance $R_N$. Different symbols refer to different crystals, as indicated in the label. Note that, despite the difference in $T_c$ among the different crystals, the gap amplitudes approximately follow a common trend. (b) Values of the gap ratio $2\Delta / k_B T_c$ as a function of the contact resistance. As in (a), different symbols refer to different crystals. Both in (a) and (b), lines are only guides to the eye.} \label{fig:gaps}
\end{figure}

Similar behaviour was observed in all the contacts showing dips. Fig.~\ref{fig:gaps} summarizes the gap values obtained from the fit of the low-temperature curves of various contacts with different resistance, in different crystals. The data show that the resistance of the contact has a very clear effect on the value of the fitting parameter $\Delta$. In contacts with low resistance, where the dips interfere with the Andreev structures (as in Fig. \ref{fig:Tdep}a) the value of $\Delta$ that results from the fit is not a good measure of the superconducting gap amplitude. Hence, the large variability in the values of $\Delta$ for $R_N < 50 \, \Omega$ simply reflects the fact that $\Delta$ is sensitive to \emph{extrinsic} effects (e.g. the Maxwell term in the contact resistance) that depend on the contact and not only on the material under study. For $R_N > 50 \,\Omega$ the number of points is much smaller, because of the aforementioned difficulty to keep the contact resistance sufficiently high. However, on increasing $R_N$ the values of $\Delta$ follow a clear trend with a tendency to saturate at 1.6-1.7 meV at the highest values of contact resistance. Note that the data points shown in Fig.\ref{fig:gaps}a come from different crystals (see the legend) that have slightly different critical temperatures: $8.48 \pm 0.02$ K for B/2, $8.86 \pm 0.07$ K for B/3,  and $9.27 \pm 0.02$ K for F/3. Despite these differences, all the values of $\Delta$ fit in a common trend. Figure \ref{fig:gaps}b reports the values of the gap ratio $2\Delta /k_BT_c$ as a function of $R_N$. The use of the gap ratio allows a safe comparison of the results obtained in different crystals. A saturating trend very similar to that of $\Delta$ is clearly seen. At high values of $R_N$, where the conditions for ballistic conduction are definitely fulfilled, the gap ratio $2\Delta /k_BT_c$ tends to the value 4.2-4.3.

\section{Conclusions}

In summary, we have reported the successful growth of Mo$_3$Al$_2$C single crystals under high-pressure, high-temperature conditions. Such synthesis conditions allow us to overcome the problems related to the low reactivity of C and the greatly different melting temperatures of the elements. Our results suggest that C-deficient stoichiometries can be achieved in Mo$_3$Al$_2$C under high pressure conditions. The current work could provide a helpful guidance for further experimental synthesis of various C-based intermetallics under high pressure.

Single-crystal X-ray refinement confirmed the high structural perfection of the grown crystals. It is remarkable that the Flack parameter values refined for all the measured crystals by using a $P4_132$ space-group model were reproducibly close to either 0 or 1, hence strongly suggesting that the studied crystals correspond to the two enantiomorphic space groups $P4_132$ and  $P4_332$.

The superconducting critical temperature of the crystals was measured by means of magnetization and transport measurements. A certain variability in $T_c$ between 8.6 K and 9.3 K was observed, depending on the growth protocol and probably related to the actual C content. The transition turned out to be very narrow, displaying $\delta T_c \simeq 0.3$ K. Some features in the resistivity (i.e., a clear minimum or a change in slope) were observed around 130 K, possibly associated with the opening of a charge density wave gap. Extensive point-contact Andreev reflection spectroscopy studies confirmed that Mo$_3$Al$_2$C is a moderately strong-coupling superconductor with $2\Delta/k_BT_c \simeq 4$. The order parameter has an $s$-wave symmetry, despite the asymmetric spin-orbit coupling arising from the lack of an inversion symmetry.

\section*{Acknowledgments}
We thank P. Moll and B. Batlogg for collaboration on early stage of this study and for fruitful discussions. We also thank J. Hulliger and T. Shiroka for critically reading the manuscript and helpful comments.

\end{document}